\documentclass{aastex63}

\usepackage{gensymb}
\usepackage{hyperref}
\usepackage [autostyle, english = american]{csquotes}
\MakeOuterQuote{"}

\begin{document}

\title{Using the Long Wavelength Array to Search for Cosmic Dawn}

\correspondingauthor{Christopher DiLullo}
\email{cdilullo@unm.edu}

\author[0000-0001-5944-9118]{Christopher DiLullo}
\affiliation{University of New Mexico \\ 
210 Yale Blvd. NE \\ Albuquerque, NM 87131}

\author[0000-0001-6495-7731]{Gregory B. Taylor}
\affiliation{University of New Mexico \\ 
210 Yale Blvd. NE \\ Albuquerque, NM 87131}

\author[0000-0003-1407-0141]{Jayce Dowell}
\affiliation{University of New Mexico \\ 
210 Yale Blvd. NE \\ Albuquerque, NM 87131}

\begin{abstract}
The search for the spectral signature of hydrogen from the formation of the 
first stars, known as Cosmic Dawn or First Light, is an ongoing effort around
the world. The signature should present itself as a decrease in the temperature
of the 21-cm transition relative to that of the Cosmic Microwave Background 
and is believed to reside somewhere below 100 MHz. A potential detection was
published by the Experiment to Detect the Global EoR Signal (EDGES) 
collaboration with a profile centered around 78 MHz of both unexpected 
depth and width \citep{bowman2018}. If validated, this detection will have 
profound impacts on the current paradigm of structure formation within $\Lambda$CDM
cosmology. We present an attempt to detect the spectral signature reported by 
the EDGES collaboration with  the Long Wavelength Array station located on the 
Sevilleta National Wildlife Refuge in New Mexico, USA (LWA-SV). LWA-SV differs from
other instruments in that it is a 256 element antenna array and offers beamforming 
capabilities that should help with calibration and detection. We report first limits from 
LWA-SV and look toward future plans to improve these limits.
\end{abstract}

\section{Introduction} \label{Intro}
The spin-flip transition in neutral hydrogen provides the opportunity to study 
the ionization history of the Universe via its corresponding 21-cm emission. 
This transition occurs when an electron in the 1S ground state changes from having
its spin parallel to that of the proton to antiparallel. The small energy
difference between these configurations corresponds to a photon with
a wavelength of 21.1 cm, or a frequency of 1420 MHz. This provides a useful probe
for observing the early Universe where neutral hydrogen is 
abundant. See \citet{furlanetto2006review,morales2010,pritchard2012} for an 
in-depth review of cosmology using the 21-cm line.

The formation of the first stars, known as Cosmic Dawn, marks the beginning of a 
phase transition for the Universe where it changes from being predominately neutral 
beforehand to predominately ionized after the Epoch of Reionization (EoR).
Tracing the relative brightness of the 21-cm emission, with respect to the 
Cosmic Microwave Background (CMB) as a function of redshift, gives insight 
into which processes dominate the excitation state of the hydrogen during times when the first
structures are forming in the Universe. The differential brightness temperature 
of the 21-cm signal relative to the CMB, $\delta T_B$, can be shown to follow:
\begin{equation}
    \label{eq:spinTemp}
    \delta T_B \approx 27 \cdot x_{HI} \sqrt{\frac{1+z}{10}} \left(\frac{T_S - T_{CMB}}{T_S}\right) \rm mK ,
\end{equation}
where $T_S$ is the spin temperature of the hydrogen, $T_{CMB}$ is the temperature of the CMB,
and $x_{HI}$ is the fraction of neutral hydrogen \citep{pritchard2012}. These quantities are all
highly dependent on redshift, $z$. We have ignored terms dealing with spatial density fluctuations.

The driving factor that determines the detectibility of the 21-cm transition, 
either as absorption or emission relative to the CMB, at a given redshift is 
the spin temperature of the hydrogen, which describes the 
relative population difference between the parallel and antiparallel spin states.
The spin temperature is affected by three main processes: absoprtion and 
emission of 21-cm photons, collisions with other hydrogen atoms and with free 
electrons, and resonant scattering of Ly$\alpha$ photons that can induce a spin-flip,
known as the Wouthuysen-Field effect \citep{wouthuysen1952,field1958}.

It is expected that the ultraviolet radiation from the first stars couples to 
the surrounding neutral hydrogen via the Wouthuysen-Field effect. This decouples 
the HI spin temperature from the temperature of the CMB and instead couples it to the colder 
kinetic temperature of the gas. The kinetic temperature of the gas, $T_K$, is colder than that of the
CMB, $T_{\gamma}$, since $T_K = T_0 (1+z)$ for a non-relativistic gas, but $T_\gamma = T_0 (1+z)^2$ for
photons, where $T_0$ is the respective temperature of each measured in the current epoch.
This would drive the spin temperature to be less than $T_{CMB}$ and so, from Equation \ref{eq:spinTemp}, 
the 21-cm signal is expected to show in absorption against the CMB for these redshifts. 
Astrophysical properties of the early Universe correlated with the features of the 21-cm signal
suggest that the absorption feature should be present at frequencies less than 100 MHz \citep{cohen2017}.
While most experiments have historically focused on detecting the later EoR emission signal,
experiments around the world are beginning to look for this absorption
feature as well \citep[e.g.][]{sokolowski2015,singh2018,price2018}. These experiments rely on using
a small number of elements to observe the entire sky and return the
sky-averaged spectrum.

After the first galaxies begin to form, their UV flux ionizes regions of the surrounding
medium. This heating drives the 21-cm signal into emission and the signal becomes 
dominated by the filling fraction of HII regions \citep{furlanetto2004}. Experiments 
attempting to measure the three-dimensional k-space power spectrum of the 21-cm line at 
these redshifts are searching for this emission signal \citep[e.g.][]{parsons2010,harker2010,deboer2017}.

The Experiment to Detect the Global EoR Signal (EDGES) collaboration has
published a potential detection of the absorption signal from Cosmic Dawn \citep{bowman2018}.
They have reported a flattened Gaussian profile centered at 78.1 MHz with a width
of 18.7 MHz and an amplitude of 0.53 K. Both the shape and amplitude of this profile 
is unexpected from conventional models \citep{furlanetto2006} and, if validated, could imply 
interactions between baryons and dark matter
\citep{barkana2018,munoz2018,berlin2018} or the presence 
of an unaccounted component in the radio background at these redshifts 
\citep{feng2018,mirocha2018,dowell2018}. There has also been much debate in the validity
of this profile \citep{hills2018,bradley2019,singh2019,sims2019}. This debate, coupled with
the wide ranging implications of this potential detection, warrants a follow up with 
a different instrument in order to validate the reported profile parameters. 
The Long Wavelength Array (LWA) offers an opportunity for follow up as the
profile center of 78.1 MHz falls within its observable band of 10-88 MHz.
The full LWA currently consists of 3 stations: LWA1, which is colocated with the Karl J. Jansky 
Very Large Array in New Mexico, USA; LWA-SV, which is located on the Sevilleta National Wildlife Refuge
in New Mexico, USA; and LWA-OVRO, which is located at the Owens Valley Radio Observatory in the Owens Valley,
California, USA. The constituent stations can operate independently and offer a beamforming mode which is described
in this paper.

The work detailed in this paper uses LWA-SV, which is the second station of the larger Long Wavelength Array.
The beamforming mode of LWA-SV offers a new method for detecting the potential Cosmic Dawn signal.
As mentioned above, other experiments that are searching for this signal observe the entire sky and generate
sky-averaged spectra. This can create challenges as sources of contamination, namely 
extragalactic point sources and galactic synchrotron and free-free emission, can obscure the signal. 
The beamforming capability of LWA-SV  allows for spatial selection on the sky in order to avoid
sources of contamination, such as 
bright sources and the Galactic plane. This will not fully remove contamination, since the beam 
side lobes will pick up some unwanted signal, but it should improve overall performance. 
Beamforming also allows us to try a different method for calibrating the array, namely 
simultaneously observing a bright calibrator source with a second beam. This allows for
\textit{in situ} astronomical calibration of the entire system instead of relying on 
laboratory measurements like other experiments.

\section{The Long Wavelength Array} \label{LWA}
LWA-SV \citep{taylor2012,cranmer2017} is an antenna array consisting of 256 dual-polarization
antennas which are arranged in a pseudo-random layout and observe within the frequency range
3-88 MHz. It can support two simultaneous beam pointings with each beam containing two tunings 
with 9.8 MHz of bandwidth\footnote{LWA-SV has been upgraded since this work was completed. It now
supports three simultaneous beams with each beam containing two tunings with 19.6 MHz of bandwidth.}. 
The array is roughly $110 \ \rm{m} \times 100 \ \rm{m}$ in the N/S 
and E/W directions, respectively.

The system must be stable in time in order to integrate the data long enough
to achieve significant signal to noise. If the system is not stable, fluctuations
in the data can obscure the signal as the data is averaged over times longer
than the fluctuation timescale. Single element experiments suffer from the need
for integration times on the order of hours. However, LWA-SV benefits from
its large effective area and should be able to reach a residual r.m.s of 50 mK, one
tenth of the signal amplitude reported by the EDGES collaboration, within
an integration time of 10's of seconds and so the electronics are not required to
be stable over large fractions of a day or over multiple days. We estimate this using
the radiometer equation within the Rayleigh-Jeans limit to relate integration time
to an observed brightness temperature via:
\begin{equation}
    \label{eq:integration}
    \Delta t = \frac{1}{2 \Delta \nu}\left[\frac{c^2 T_{sys}}{\nu^2 T_{rms} A_e} \right]^2 ,
\end{equation}
where $\Delta \nu$ is the channel bandwidth, $\nu$ is the tuning frequency, $A_e$ is the 
effective area of the array, and $T_{sys}$ and $T_{rms}$ are the system and r.m.s. brightness temperatures,
respectively. Assumed values\footnote{\url{http://lwa.phys.unm.edu/obsstatus/obsstatus006.html\#toc14}}
for a few frequencies and results for the integration time are summarized in Table \ref{table:radiometry}.

\begin{deluxetable}{cccc|c}
\tablecolumns{6}
\tablewidth{0pt}
\tablecaption{Radiometer Equation Assumptions and Results for $T_{rms} = 50 \ \rm{mK}$.}\label{table:radiometry}
\tablehead{\colhead{$\nu$} & \colhead{$\Delta \nu$} & \colhead{$A_e$} & \multicolumn{1}{c|}{$T_{sys}$} & \colhead{$\Delta t$}}
\startdata
55 MHz & 9.57 kHz & $1900 \ \rm{m}^2$ & 3840 K & 73.4 s \\
65 MHz & 9.57 kHz & $1360 \ \rm{m}^2$ & 2500 K & 32.0 s \\
75 MHz & 9.57 kHz & $1020 \ \rm{m}^2$ & 1740 K & 15.5 s \\
\enddata
\end{deluxetable}

The signal chain of LWA-SV is straight forward, but there are many places that can
introduce instabilities. The entire chain can be broken into the front end electronics (FEEs)
and the back end electronics. The FEEs are dual polarization receiver boards located on
each of the dipoles. They provide initial amplification and low-pass filtering of the signal before it is passed
to the back end electronics. The back end of LWA-SV consists of two components,
the analogue signal processor (ASP) and the advanced digital processor (ADP). 
The ASP electronics apply a second gain stage to the dual polarization analogue signal from each dipole
and apply a bandpass to the signal. These signals are then passed along to ROACH-2 boards 
within ADP which digitize them and compute the Fast Fourier Transform (FFT) in order to 
output a complex spectrum. These data are then passed into the beamformer, after which the full
beamformed data is recorded onto disk. 

To ensure that the data remains unobscured by any fluctuations in the electronics
chain of LWA-SV, we verified that an unstructured noise signal injected into the back end
electronics integrated as expected. A Noisecom, Inc. model 
3201K\footnote{\url{https://www.noisecom.com/products/calibrated-sources/nc3200-coaxial-noise-sources}} noise source was 
connected to one input of ASP and all other inputs were zero weighted. The choice of which ASP input channel the noise source is
injected into should not affect the results. The ASP electronics are designed to have channel independent outputs
and this has been verified through lab measurements\footnote{See \href{http://www.phys.unm.edu/~lwa/memos/memo/lwa0201.pdf}{LWA Memo \#201}}.
If the station is generally stable over a given length of time,
the data output by the system for a noise-like input should integrate with an r.m.s. which goes like:
\begin{equation}
    \label{eq:rms}
    \sigma \propto t^{-1/2},
\end{equation}
since this is true for Gaussian data. It is important to note that this only tests the 
stability of the back end electronics and does not account for potential instabilities 
in the FEE.

\begin{figure}
    \centering
    \includegraphics[width=\textwidth]{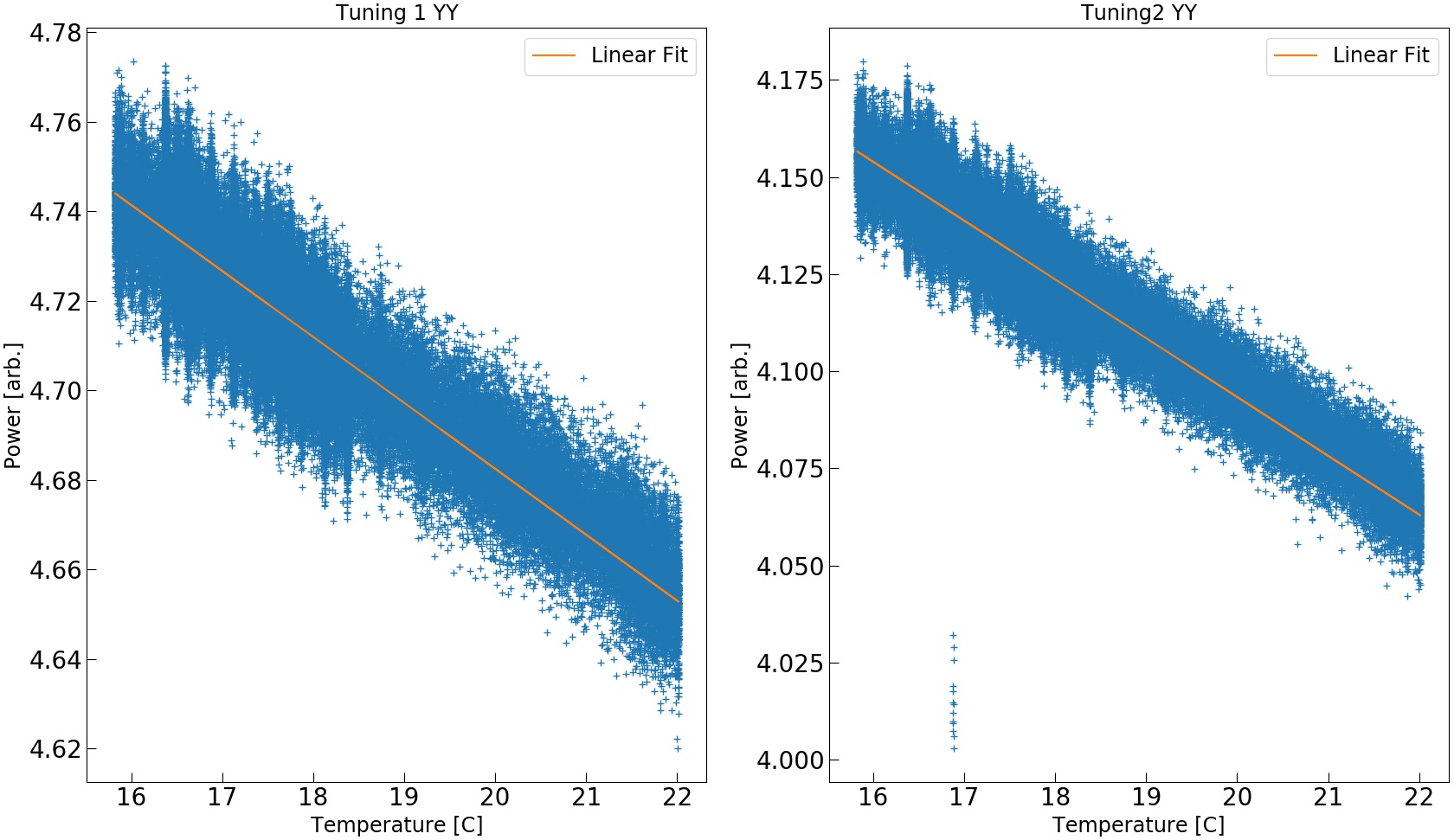}
    \caption{Median power across frequency vs ASP electronics temperature. 
            The linear fit is used to detrend air conditioning cycle effects.}
    \label{fig:acFits}
\end{figure}

\begin{figure}
    \centering
    \includegraphics[width=\textwidth]{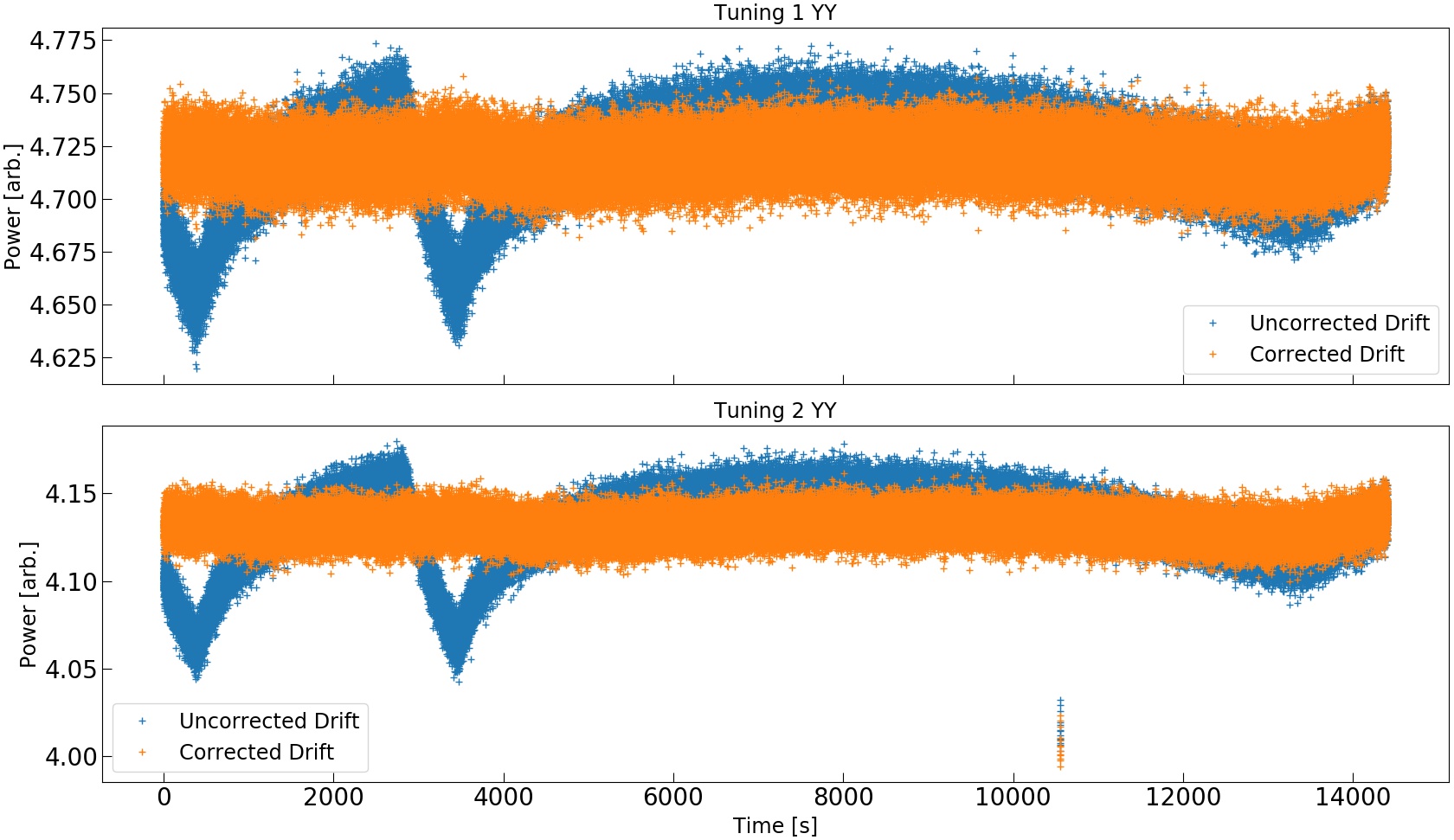}
    \caption{Uncorrected and corrected median drifts. The uncorrected drifts have large temporal structure in them as a result
             of the air conditioning cycle within the electronics shelter of LWA-SV. This causes variations in the response of the ASP electronics which induce power variations. The corrected drifts have been detrended and are centered about the mean of the uncorrected drifts.}
    \label{fig:acDrifts}
\end{figure}

Temperature variations in the electronics shelter, due to the air conditioning cycle, cause
the amplifier responses in the electronics to vary creating power variations in the data.
We correct for these variations by fitting a linear relation between the observed median drift and 
the median temperature of the ASP electronics which is used to detrend the observed drift. The detrended
drifts are then multiplied by the mean power of the original drift. The linear fit and drifts, before and after correction,
are shown in Figures \ref{fig:acFits} and \ref{fig:acDrifts}, respectively.

\begin{figure}
    \centering
    \includegraphics[width=\textwidth]{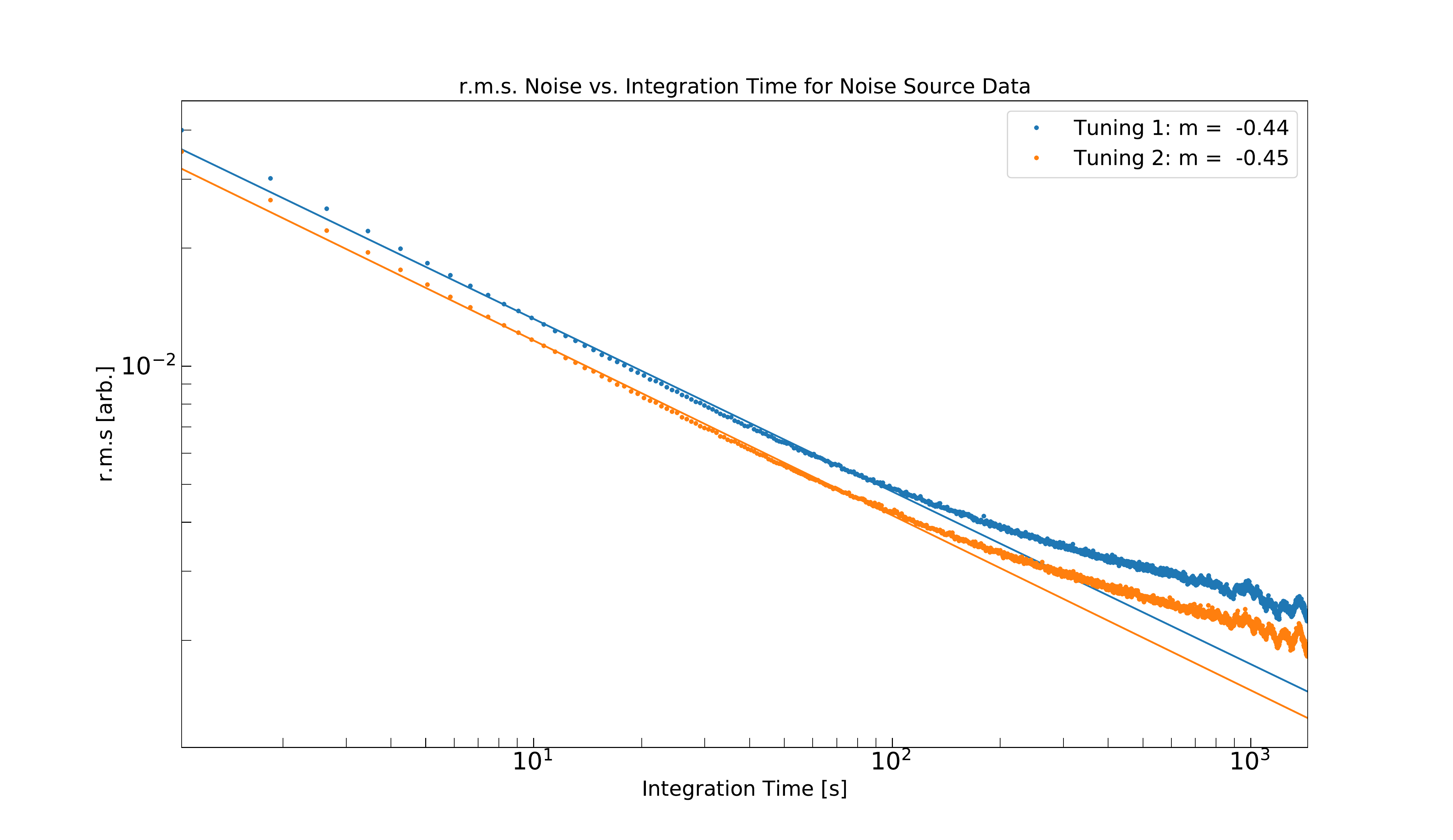}
    \caption{Log-log plot of r.m.s. noise vs integration time for a Gaussian noise 
             source. Unstructured Gaussian data should integrate down with a slope of -0.5.
             The linear fits are computed over times $t \leq 120 \ \rm{s}$.}
    \label{fig:NoiseSource}
\end{figure}

After correcting for power variations induced by the air conditioning cycle, we iteratively compute the r.m.s.
noise of the data at different integration times to test overall system performance. In order to test how the
data as a whole integrates, we use a bootstrapping method to randomly select 75\% of frequency channels at each
iteration. We then compute the r.m.s. in each of these channels and report the mean r.m.s. noise for that integration time.
The results are shown in Figure \ref{fig:NoiseSource} for integration times up to 24 minutes. We fit a line using the data
for integration times of $t \leq 120 \ \rm{s}$ and report the best fit slopes. It is apparent that while we do not integrate
down with the ideal slope of -0.5 (see Equation \ref{eq:rms}), we have good performance with slopes that are $\sim 0.05$ away
from ideal. It is unclear why we begin to plateau after integration times larger than a few minutes. The larger variations
in the curves at the largest integration times are likely statistical in origin. The uncertainty in the r.m.s. at these
integrations is dominated by statistical uncertainty arising from small sampling size. A 4 hour observation will only have 10
data points with which to compute the r.m.s. for an integration time of 24 minutes. We conclude that the system is stable
at least on the order of a few minutes. This should be enough to detect the Cosmic Dawn signal (see Table \ref{table:radiometry}).

\section{Observations} \label{Obs}
The LWA1 Low Frequency Sky Survery \citep[LFSS;][]{dowell2017} covered the entire sky visible to
the LWA and was used to identify the coldest region on the sky. The reasoning for this was to
minimize foreground contamination. The coldest region on the sky should offer the best chance
to disentangle the Cosmic Dawn signal from galactic foregrounds. The cold region that was 
identified and observed has J2000 coordinates of a RA of $9^{\rm h} \ 38 ^{\rm m} \ 40.56 ^{\rm s}$ and a DEC 
of $+ 30^{\degree} \ 49' \ 1.4 ''$, hereafter referred to as the "Science Field". In order to 
calibrate the observations, we simultaneously point a second beam at Virgo A, located at J2000 coordinates of a 
RA of $12^{\rm h} \ 30 ^{\rm m} \ 49.42 ^{\rm s}$ and a DEC of 
$+ 12^{\degree} \ 23' \ 28 ''$. This allows for simultaneous \textit{in situ} calibration 
using a source which $\sim 43.7\degree$ away on the sky. This helps
calibrate out time dependent systematics. The angular separation between Virgo A
and the Science Field is large which means that this calibration will not accurately account for ionospheric 
effects since the two beams will each suffer unique perturbations as the signal paths through the ionosphere are different.
This can be addressed with an ionospheric model, but that will considered in the future.
However, Virgo A is the closest object which is also significantly bright enough to use for calibration.

Observations were taken on 2019, October $11^{\rm th}$ for 3 hours beginning at 15:58:00 UTC.
This time range captures both Virgo A and the Science Field at high elevation, with their
elevations being similar during the midpoint of the observation. This minimizes
any elevation dependent effects during the middle of the observation. Both beams contain two tunings
centered at 67 and 75 MHz, each with a bandwidth of 9.8 MHz. Accounting for rolloff at the edges of the band,
which limits us to the inner 80\%, this yields almost continuous coverage between 63 and 79 MHz. 
There is a small gap in coverage with a width of about 0.5 MHz around 71 MHz.

We chose LWA-SV's spectrometer mode\footnote{See \href{http://www.phys.unm.edu/~lwa/memos/memo/lwa0177c.pdf}
{LWA Memo \#177}.}, which channelizes the data and averages individual integrations to ouput 
time averaged spectra, for data acquisition. Setting the number of frequency channels to 1024 and the number of integrations 
per spectrum to 768 provided us with spectra with 9.56 kHz frequency resolution and 80 ms time resolution.
We obtained spectra for the two linear polarizations XX and YY, which are elongated in the E/W and N/S directions,
respectively. All data reduction was done using modules in the LWA Software 
Library\footnote{\url{http://fornax.phys.unm.edu/lwa/trac/wiki}}\citep[LSL;][]{dowell2012LSL}.

We flag RFI using a pseudo-spectral kurtosis flagging criterion \citep{nita2010statistics,nita2010generalized} that flags 
data with a spectral kurtosis outside of $3\textrm{-}\sigma$ from the mean. The spectral kurtosis is "pseudo" because the spectra that 
is output by the spectrometer mode is an average over our chosen number of 768 integrations per spectrum. We also use a smooth-bandpass
model, created by smoothing the data along the frequency axis, to flag any frequency channels with an average power greater than $15\textrm{-}\sigma$ 
from the mean. This helps flag narrow-band RFI arising from digital television pilot tones that fall within the observed band that are not captured by the 
pseudo-spectral kurtosis criterion. One RFI source was identified as originating from a noisy electrical pole located roughly a mile from the station,
but it is unclear what effect this has on the overall quality of the data.

The spectra output by LWA-SV are initially in units of power on an arbitrary scale. It
is therefore necessary to calibrate the data in order to yield a temperature spectrum. 
The calibration using Virgo A works by simulating LWA-SV's beam pattern at a given frequency
on Virgo A at the midpoint of the observation. This beam pattern is then used in conjunction with the Global Sky 
Model \citep[GSM;][]{deOliveira2008} to obtain a simulated measurement of the temperature of Virgo A at that
frequency. This temperature is then divided by the arbitrary power reported by LWA-SV to generate a scaling
coefficient in units of [K / Power] for that frequency. This is carried out on a per frequency channel basis, in 
order to get a set of scaling coefficients that can be applied to the spectrum of the Science Field. 
These scaling coefficients are expected to be independent of the spectral structure of Virgo A and should maintain 
the spectral shape of the Science Field.

The accuracy of our beam models for each pointing on the sky and the accuracy of the GSM temperatures for Virgo A 
can obscure the calibration and induce spectral structure in the calibrated spectra. 
The beam model used incorporates an electromagnetic simulation of the beam pattern of a single dipole and
models of the responses of the electronics along the signal path. However, it is important to note that the accuracy of the beam model
is extremely difficult to verify as measuring the pattern of the entire station is a major challenge. The accuracy of the dipole simulation
is also difficult to measure in the field. Possible methods of measuring the station beam pattern are discussed in Section \ref{beamforming}.
The accuracy of the GSM is dependent on frequency and position on the sky. However, at these low frequencies and Galactic latitudes, the
authors report the accuracy being $\sim 10\%$ or better with respect to the input maps. \citet{dowell2017} also found a similar level
of agreement between a 74 MHz map made with the first station of the LWA (LWA1) and the GSM in the region of the sky near Virgo A.

\section{Results and Current Limits} \label{Results}
\begin{figure}
    \centering
    \includegraphics[width=\textwidth]{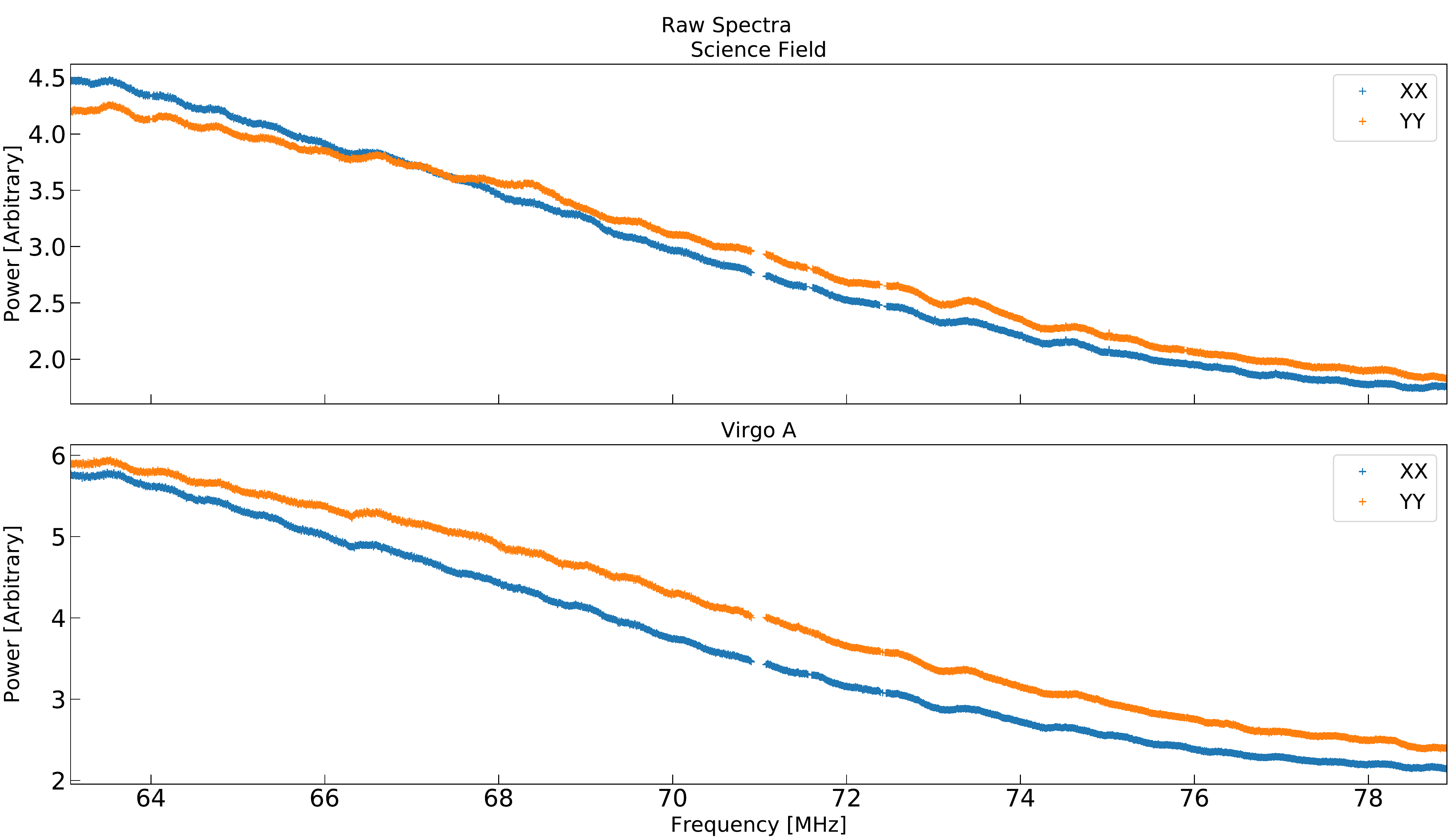}
    \caption{Raw spectra of the Science Field (top) and Virgo A (bottom) for XX and YY polarizations. These spectra
            consist of 2 minutes of integrated data.}        

    \label{fig:rawSpectra}
\end{figure}

\begin{figure}
    \centering
    \includegraphics[width=\textwidth]{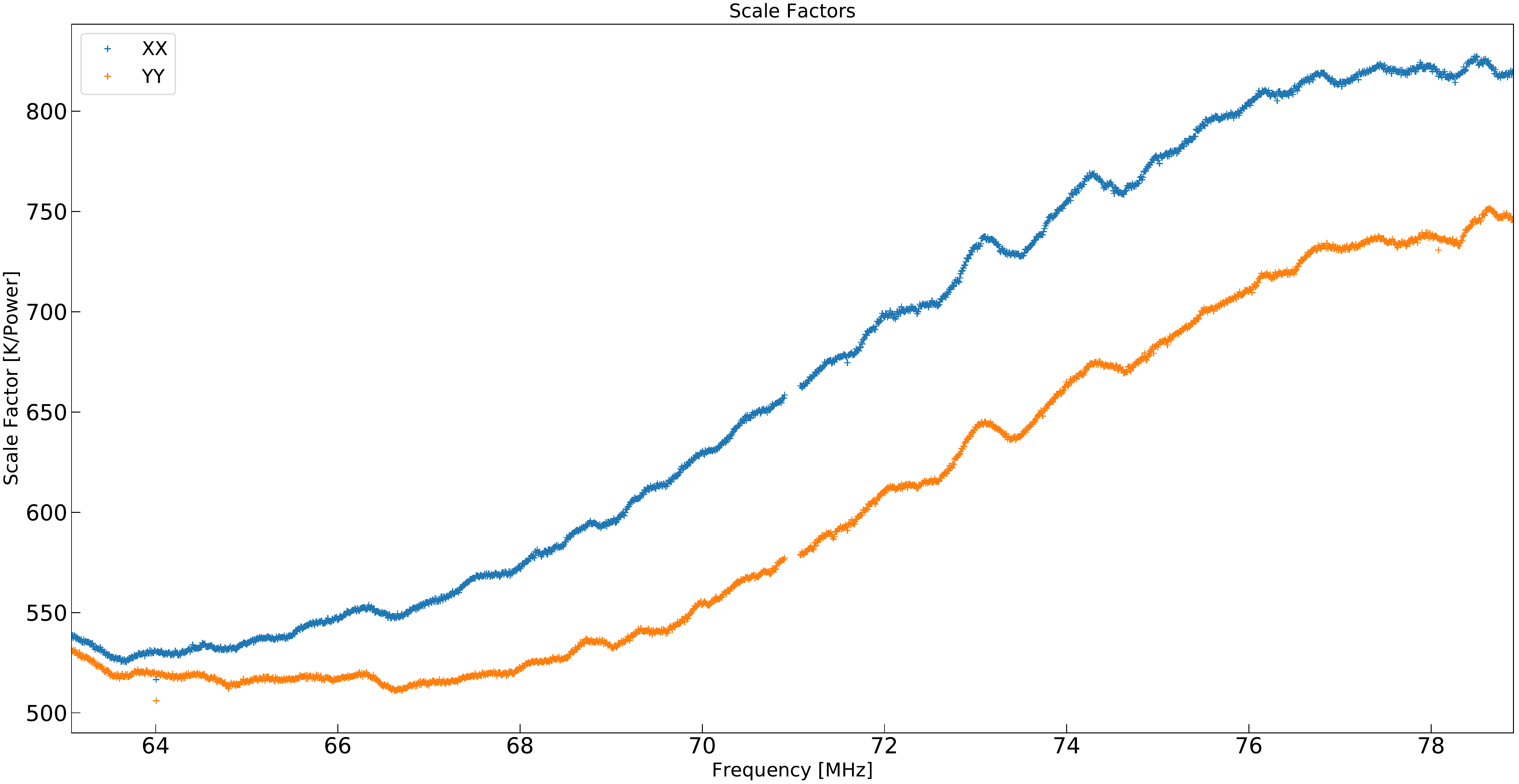}
    \caption{Temperature scale factors derived from the observed power of Virgo A
            and simulated beam temperatures from the Global Sky Model.}
    \label{fig:ScaleFactors}
\end{figure}

The observed uncalibrated spectra of the Science Field and Virgo A are presented
in Figure \ref{fig:rawSpectra}. These are the median spectra of 2 minutes of data 
selected from the middle of the observation described in Section \ref{Obs}. It is 
apparent that the low brightness Science Field suffers from more spectral structure
than the brighter Virgo A field. The scale factors derived from our calibration scheme,
described in Section \ref{Obs}, are presented in Figure \ref{fig:ScaleFactors}. 
These factors should help calibrate structures, whether celestial, ionospheric, or system-based 
in nature, and should smooth the spectra of the Science Field.

It is important to note that the scale factors do have spectral structure that we believe
should not be present. The scaling should be spectrally smooth in nature,
but it is clear from Figure \ref{fig:ScaleFactors} that this is not the case as many bumps are
present. The source of this structure is believed to be mostly related to the chromaticity
of the beam of LWA-SV. Potential sources of spectral structure which are not captured in the
current temperature calibration scheme, such as beam chromaticity, will be discussed in Section
\ref{discussion}.

\begin{figure}
    \centering
    \includegraphics[width=\textwidth]{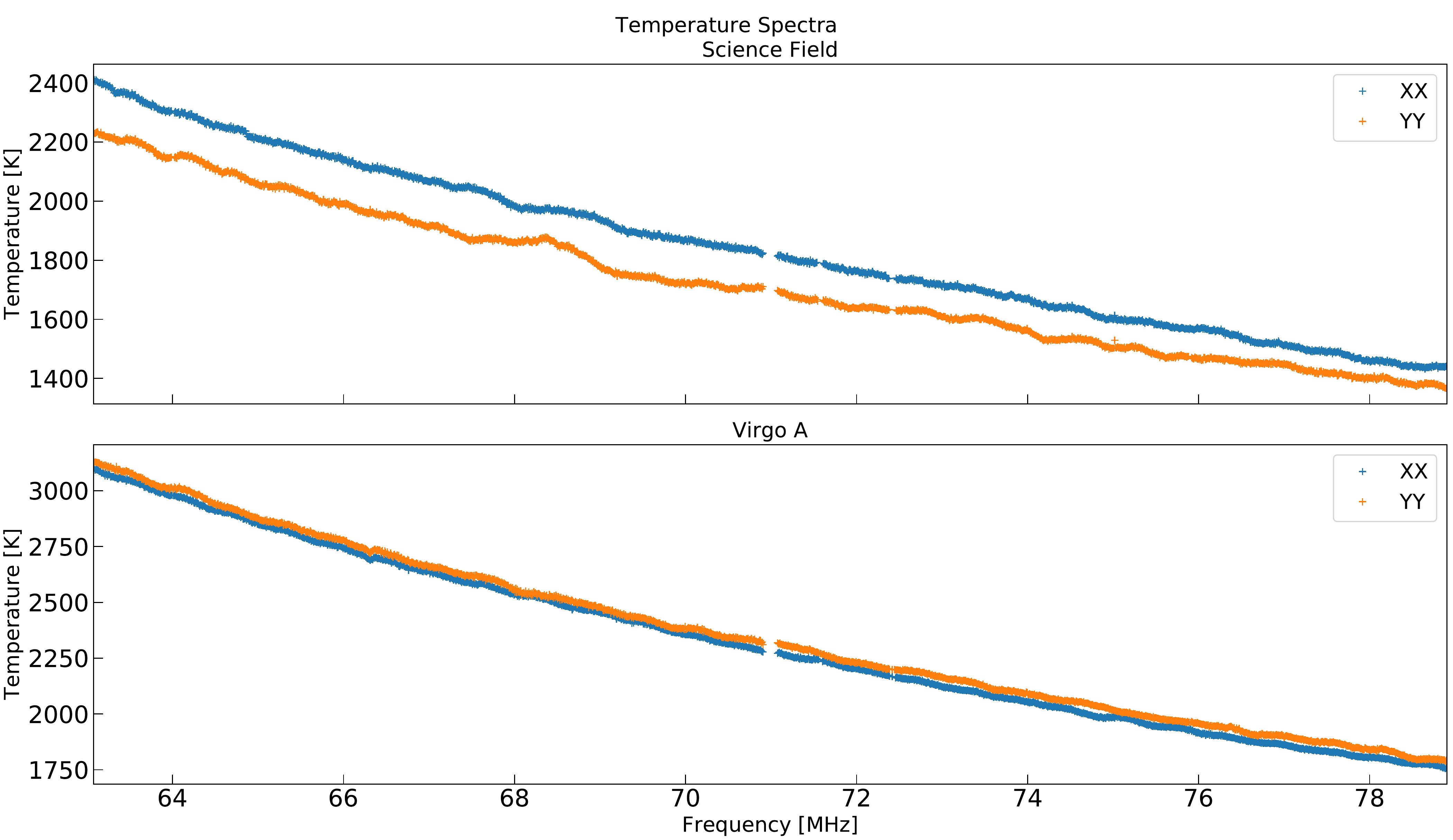}
    \caption{Calibrated spectra in units of Kelvin. These are generated by multiplying the spectra shown in 
            Figure \ref{fig:rawSpectra} with the curves in Figure \ref{fig:ScaleFactors}.}
    \label{fig:tempSpec}
\end{figure}

The temperature calibrated Science Field and Virgo A spectra are presented in Figure \ref{fig:tempSpec}.
The spectra are generally smooth in nature with some small structure present in the Science Field spectra.
These unsmooth features will not be captured by a smooth polynomial fit and will remain present in the 
residuals. 

\begin{figure}
    \centering
    \includegraphics[width=\textwidth]{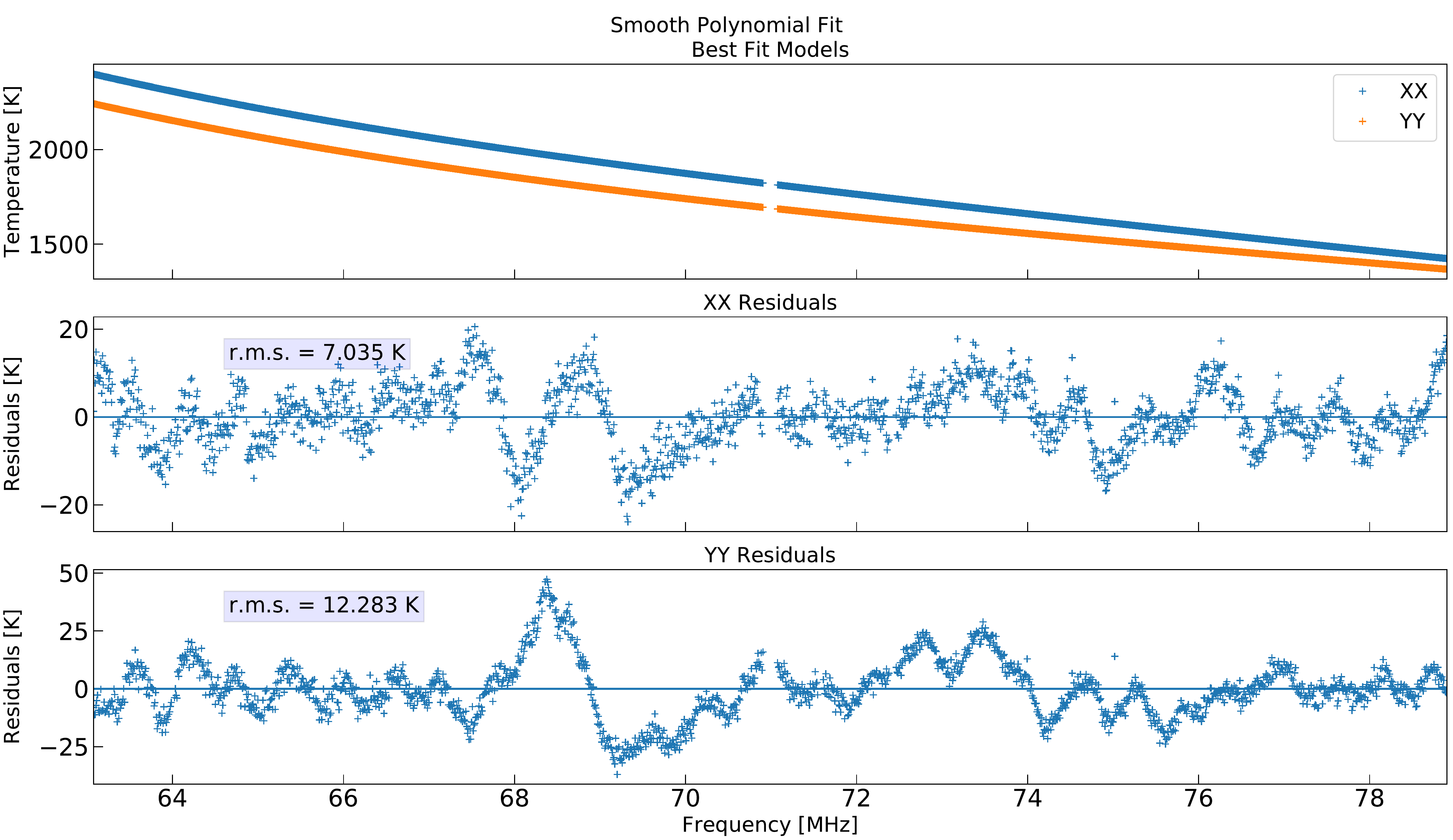}
    \caption{Best fit N=5 smooth polynomial model for each polarization. Residuals are shown in the bottom plots 
            with zero marked with a horizontal line and r.m.s. reported in the blue box.}
    \label{fig:SmoothPolyModel}
\end{figure}

\begin{figure}
    \centering
    \includegraphics[width=\textwidth]{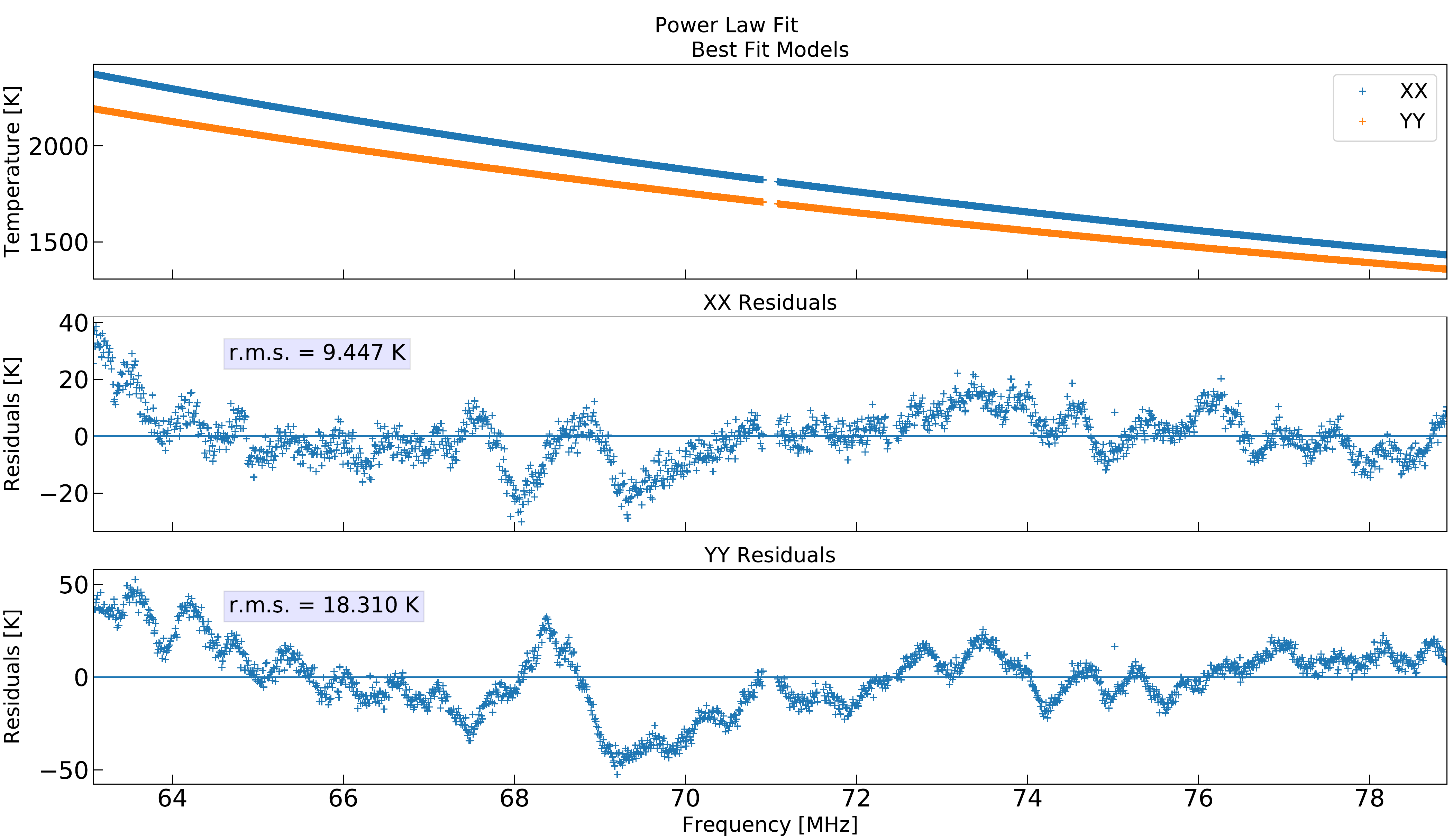}
    \caption{Best fit power law model for each polarization. Residuals are shown in the bottom plots with zero marked
            with a horizontal line and r.m.s. reported in the blue box.}
    \label{fig:PowerLawModel}
\end{figure}

We investigate the performance of two foreground models of differing complexity in order to determine 
the extent of model dependence in the r.m.s. of the residuals. The first model we investigate is a simple
power law of the form:
\begin{equation}
    \label{eq:logModel}
    T(\nu) = k \left(\frac{\nu}{\nu_0}\right)^\alpha,
\end{equation}
where $k$ is a proportionality constant, $\nu_0$ is a reference frequency, and $\alpha$ is the spectral index.
The second model is a smooth 5-term polynomial of the form:
\begin{equation}
    \label{eq:skyModel}
    T(\nu) = \sum_{n=0}^{4} a_{n} \left(\frac{\nu}{\nu_0}\right)^{n-2.5},
\end{equation}
where $\nu_0$ is a reference frequency. We chose $\nu_0$ to be the center frequency of the band for
both models. These functions should capture the overall smooth shape of the spectrum, which is expected in
this frequency regime. The best fit parameters for both models are presented in Table \ref{table:ForegroundModels}.
The best fit models, with residuals, are presented in Figures \ref{fig:SmoothPolyModel} and \ref{fig:PowerLawModel}.

\begin{deluxetable}{cccc}
\tablewidth{0pt}
\tablecaption{\label{table:ForegroundModels} Foreground Model Best Fit Parameters}
\tablehead{\colhead{Model} & \colhead{Parameter} & \colhead{XX Polarization} & \colhead{YY Polarization}}
\startdata
                            & $a_0$  & $7.49 \times 10^4 \pm 1.43 \times 10^4$ & $2.29 \times 10^4 \pm 2.51 \times 10^4$   \\
                            & $a_1$  & $-2.69 \times 10^5 \pm 5.78 \times 10^4$ & $-5.96 \times 10^4 \pm 1.01 \times 10^5$ \\
N=5 Smooth Polynomial       & $a_2$  & $3.66 \times 10^5 \pm 8.74 \times 10^4$ & $5.21 \times 10^4 \pm 1.53 \times 10^5$   \\
                            & $a_3$  & $-2.18 \times 10^5 \pm 5.86 \times 10^4$ & $-1.10 \times 10^4 \pm 1.02 \times 10^5$ \\
                            & $a_4$  & $4.81 \times 10^4 \pm 1.47 \times 10^4$ & $-2.72 \times 10^3 \pm 2.57 \times 10^4$   \\
\hline \\
                            & $\alpha$ & $-2.26 \pm 1.89 \times 10^{-3}$  & $-2.14 \pm 3.83 \times 10^{-3}$  \\
Power-Law                   & & \\
                            & $k$      & $3.26 \pm 5.36 \times 10^{-5}$   & $3.23 \pm 1.09 \times 10^{-4}$   \\
\enddata
\end{deluxetable}

\begin{figure}
    \centering
    \includegraphics[width=\textwidth]{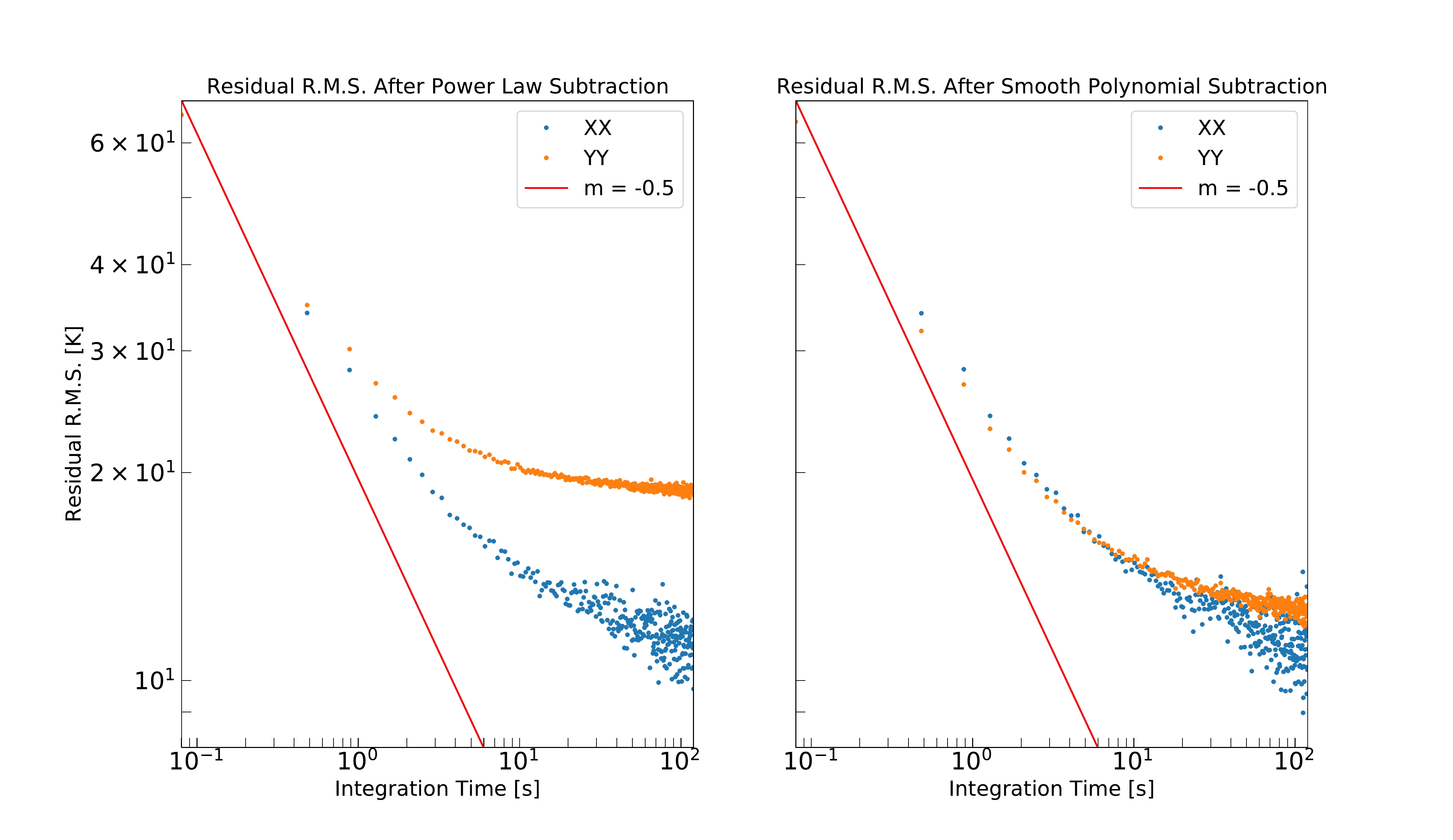}
    \caption{Average residual r.m.s. vs. integration time for each polarization. Unstructured
            data should go like $\sigma \propto t^{-1/2}$ (red solid line).}
    \label{fig:rms}
\end{figure}

We check the performance of the system by iteratively computing the average residual r.m.s. of time-averaged spectra for different integration times, ranging from the native 80 ms resolution of the observation to 2 minutes. The data corresponding to the central 20 minutes of the observation are selected and time-averaged accordingly at each iteration to produce a number of time-averaged spectra. The foreground model is then subtracted from these time-averaged spectra and a random sample of 75\% of the residual spectra are selected. We then compute the residual r.m.s. across frequencies for these residuals and compute the average residual r.m.s. across the chosen sample. This allows us to estimate the behavior of the entire distribution of residuals as a function of integration time.
We look for the r.m.s. to follow a similar trend to that seen in Figure \ref{fig:NoiseSource}.
This would mean that we have correctly accounted for all forms of spectral contamination both from the system
and from contaminating sources on the sky. The average residual r.m.s. vs. integration time for both models
is shown in Figure \ref{fig:rms}. It is apparent that the data does not integrate like unstructured data,
but rather we begin to plateau after $\sim 10 \ \rm{seconds}$ of integration. 

\

\section{Discussion} \label{discussion}
The r.m.s. of the residuals in Figures \ref{fig:SmoothPolyModel} and \ref{fig:PowerLawModel} are much higher 
than the required levels to verify the potential detection published by EDGES. In order to truly verify the claim,
we require the residual r.m.s. to be at most on the order of $50 \ \rm{mK}$. This means that our reported r.m.s. level
of $\approx10 \ \rm{K}$ for the XX polarization is two orders of magnitude above what we require. This high r.m.s.
is likely an effect of our inability to improve signal to noise (SNR) via integration times longer than
a few seconds. It is useful to break the discussion into general categories that help narrow focus onto specific
issues that are contributing to the limits reported above. The issues generally
fall into two categories: Calibration and Modelling.

\subsection{Calibration}

Figure \ref{fig:rms} highlights that something in the system is currently
prohibiting us from integrating data for long periods of time and this severely limits our SNR.
The results of our initial investigation that injected a noise source into the back end of LWA-SV,
summarized in Figure \ref{fig:NoiseSource}, seem to suggest that the issue is not the stability of
LWA-SV back end electronics. This isolates the problem to be somewhere in either the front end electronics 
or the chromaticity of the station beam.

The front end of LWA-SV can suffer from issues such as the impedance mismatch between the dipoles
and the FEE, standing waves between the dipoles and their ground screens, and mutual coupling between
individual dipoles within the array. There have been impedance measurements made on LWA dipoles \citep{hicks2012},
but these measurements are difficult to make for a single dipole and near impossible for the entire constructed
array. Nevertheless, these models are used to take the impedance mismatch into account with the assumption that all dipoles behave 
the same. Standing waves between the dipoles and their ground screen would manifest as waves within the observed spectra. 
These should stand out in the residuals and are not seen in our spectra. Mutual coupling between elements has been 
shown to cause significant differences in response for each antenna, but should average out across the entire array for 
beamformed observations \citep{ellingson2011}.

The frequency dependence of the dipole gain pattern\footnote{See LWA Memos \href{http://www.phys.unm.edu/~lwa/memos/memo/lwa0175.pdf}{\#175} 
and \href{http://www.phys.unm.edu/~lwa/memos/memo/lwa0178a.pdf}{\#178.}} 
and beam size contributes a major challenge for these 
types of experiments. This chromaticity leads to varying responses, both from the main lobe and the side lobes,
with frequency which manifest as spectral features that obscure the signal. The full width at half maximum (FWHM)
of the main lobe response can vary by $\approx0.5\degree$ across our band of $\approx 63 - 79 \ \rm{MHz}$. The total beam pattern will also 
change as a function of pointing on the sky which can cause time-dependent variations in the response as the beam 
tracks a source over the duration of an observation. Work aiming to solve the chromaticity and directionality of the
beam is presented in Section \ref{beamforming}.

\subsection{Modelling}
Modelling plays a crucial role in the work presented above both in our temperature calibration and in our subtraction
of galactic foregrounds. Our chosen temperature calibration scheme makes use of the GSM \citep{deOliveira2008} at $1\degree$ 
resolution; however, this is not the only sky model that we can use. The LFSS has been used to create the Low Frequency Sky Model
(LFSM). This would seem to be the most logical sky model to use in order to maintain consistency of instrumentation, but the LFSM
has an angular resolution of $5.1\degree$ which is currently larger than the size of the beam main lobe across our entire band.
If the beam size is sub-resolution of  the chosen sky model, then the temperatures derived for Virgo A will not accurately match
the observed power and the scaling will be in error. The LFSM should be investigated more in the future as we further develop custom
achromatic beams. 

Concerns have also been raised about the GSM using a cubic-spline interpolation between input surveys which does not take physical
constraints into account \citep{rao2017modeling}. The Global Model for the Radio Sky Spectrum \citep[GMOSS;][]{rao2016gmoss}
has been proposed as an alternative which takes physical constraints into account when interpolating between input surveys.
However, GMOSS is also only available at $5\degree$ resolution and so we have not investigated its potential as our beam size is
sub-resolution. This would lead to inaccurate temperature scaling, as stated above. Therefore, we leave the investigation into
GMOSS's performance for future work.

There is also debate surrounding how to properly model Galactic foregrounds that obscure the Cosmic Dawn
signal. \citet{bowman2018} use a physically motivated model which accounts for galactic synchrotron emission
and Earth's ionosphere. A derivation of this physically motivated model can be found in \citet{hills2018}.
However, \citet{singh2019} express concerns that radiometers that observe the entire sky can produce spectra
which will not be fit well by physically motivated polynomials. This is due to output spectra of the radiometer
being some combination of spectra of various objects across the sky that have different spectral shapes. Thus,
the output will not be well described by a single physically motivated power law. Instead, they propose the 
usage of maximally smooth functions \citep{rao2015,rao2017modeling}. These should remove
the smooth structure in the sky-averaged spectrum and reveal the Cosmic Dawn signal in the residuals; however, it
is unclear what degree polynomial is needed to properly model systematic effects without over-modelling and
washing out the Cosmic Dawn signal. It is for this reason we have looked at the performance of two different
models of varying complexity in this work.

The simple power law foreground model seems to model the observed spectrum almost as well as the $5^{\rm th}$
order smooth polynomial model. The similarities between the residuals presented in Figures \ref{fig:SmoothPolyModel}
and \ref{fig:PowerLawModel} show that most of the residual structure is on relatively small spectral scales. However,
both residual r.m.s. limits presented here are overly optimistic since we have not considered the effects of signal loss
due to over-modelling \citep{cheng2018}. This is especially true for the $5^{\rm th}$ order smooth polynomial model
as it will be highly correlated with smaller spectral scales. \citet{bernardi2015} investigated the effects of 
instrumental response over a large bandwidth and found the assumption that low order foreground models will
appropriately account for instrumental effects is too optimistic. They claim the signal should be detectable
and will not be over-modelled by a foreground polynomial of order $\approx 5$. We expect even more
on-sky structure with a beamforming approach, so a higher order polynomial model may be necessary. However, they
jointly model the foreground emission and the 21-cm signal, which will account for correlations between the model parameters,
and consider a much larger bandwidth than observed here. A 21-cm signal with spectral structure on the order of the bandwidth
observed here can lead to large correlations between the $5^{\rm th}$ order smooth polynomial foreground model and 
the 21-cm signal, leading to possibly significant signal loss. These effects will be less significant with a more simple
foreground model like the power law shown in Figure \ref{fig:PowerLawModel}. We leave a more in depth analysis of signal
loss to future work.

\section{Custom Beamforming} \label{beamforming}
\begin{figure}
    \centering
    \includegraphics[width=\textwidth]{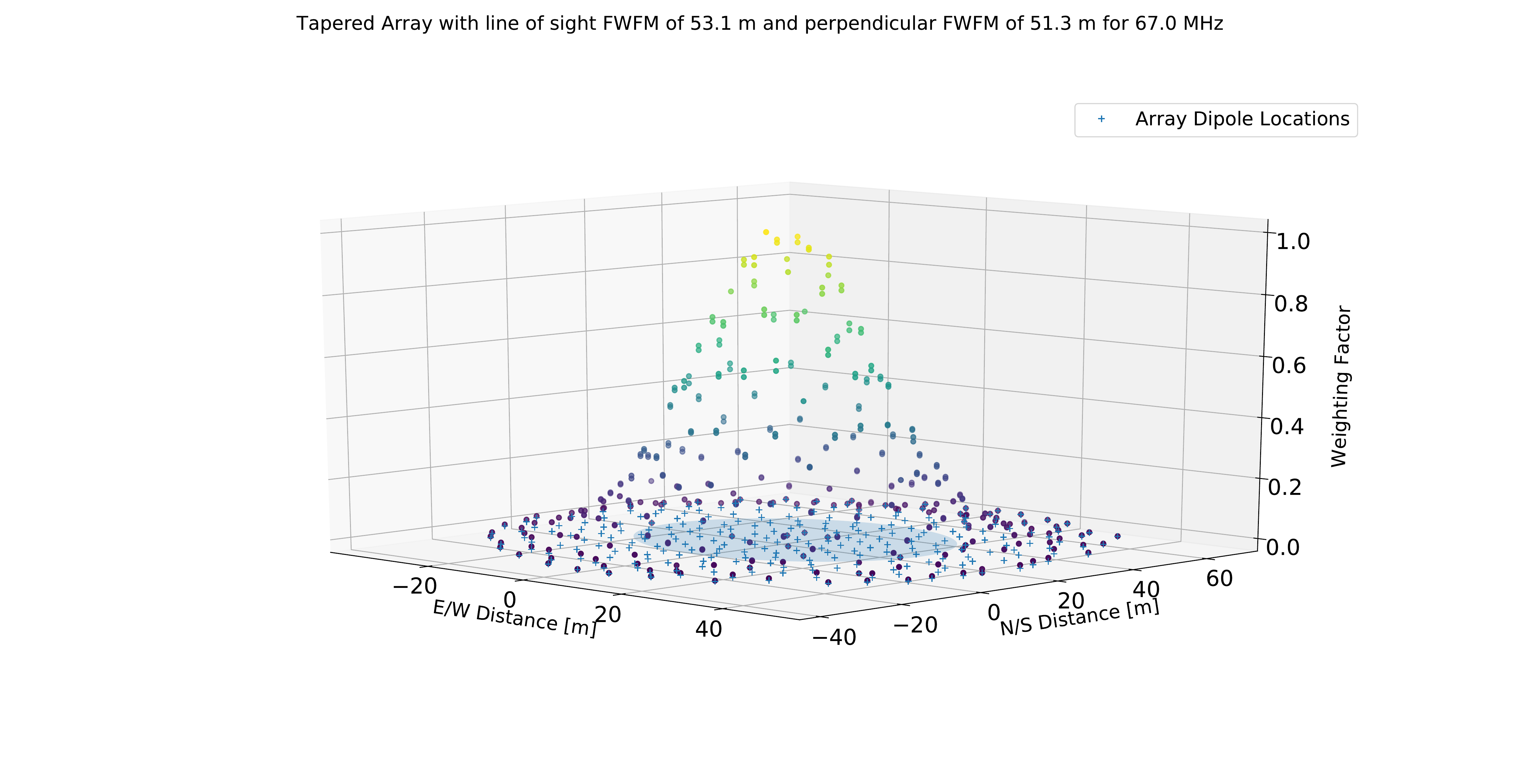}
    \caption{3-d weighting plot showing the antenna locations (crosses) and their weight values (dots).
            The weighting function is a Gaussian concentric with the array. The area of the array within the 
            full width at fifth maximum (FWFM) is shaded blue. This area has a major axis of $53.1 \rm m$ in the N/S direction
            and a minor axis of $51.3 \rm m$ in the E/W direction. This was generated for a pointing at $180\degree$ azimuth and
            $75\degree$ elevation to obtain a $5\degree$ beam at $67$ MHz.}
    \label{fig:weights}
\end{figure}

The chromaticity and directionality of the beam are two major sources of spectral
structure for our current results. The development of a framework for custom beamforming
is necessary if we wish to push our limits further into the regime were the Cosmic Dawn
signal is detectable. LWA-SV offers an advantage as an antenna array since elements can be weighted
in such a way to shape the beam into a custom configuration. This should allow for not only
custom shaped beams for some given pointing, but also achromatic beams at that pointing across
the full band. We have begun work on simulating and forming custom beams with LWA-SV. The simplest first 
case is to make a beam with a circular main lobe with some defined FWHM for a given frequency at any pointing on the sky. This can 
be achieved by weighting the array in such a way that compensates for the projection of its elliptical shape onto the sky. We begin by
defining a new coordinate system, $(x',y')$, cocentered on the array with the $x'$- and $y'$-axes corresponding to directions along
and perpendicular to the line of sight for a given azimuth, respectively. We then compute the physical diameter
that would correspond to a defined beam width via:
\begin{equation}
    D_1 = \frac{c}{\nu \theta},
\end{equation}
where c is the speed of light, $\nu$ is frequency, and $\theta$ is the user-defined beam width in radians.
A second diameter is then computed which accounts for projection effects along the line of sight, given by:
\begin{equation}
    D_2 = \frac{D_1}{\rm{sin}(e)},
\end{equation}
where $e$ is the elevation of the beam pointing center. The weighting is set by defining an ellipse that is aligned
with the $x'$- and $y'$-axes and has a major axis of length $D_2$ and minor axis of length $D_1$. We apply a Gaussian
taper from the center of the array with full widths at fifth maximum (FWFM) of $D_2$ and $D_1$ along the $x'$- and $y'$-
axes, respectively. An example of this weighting scheme can be seen in Figure \ref{fig:weights}. 

\begin{figure}
    \centering
    \includegraphics[width=\textwidth]{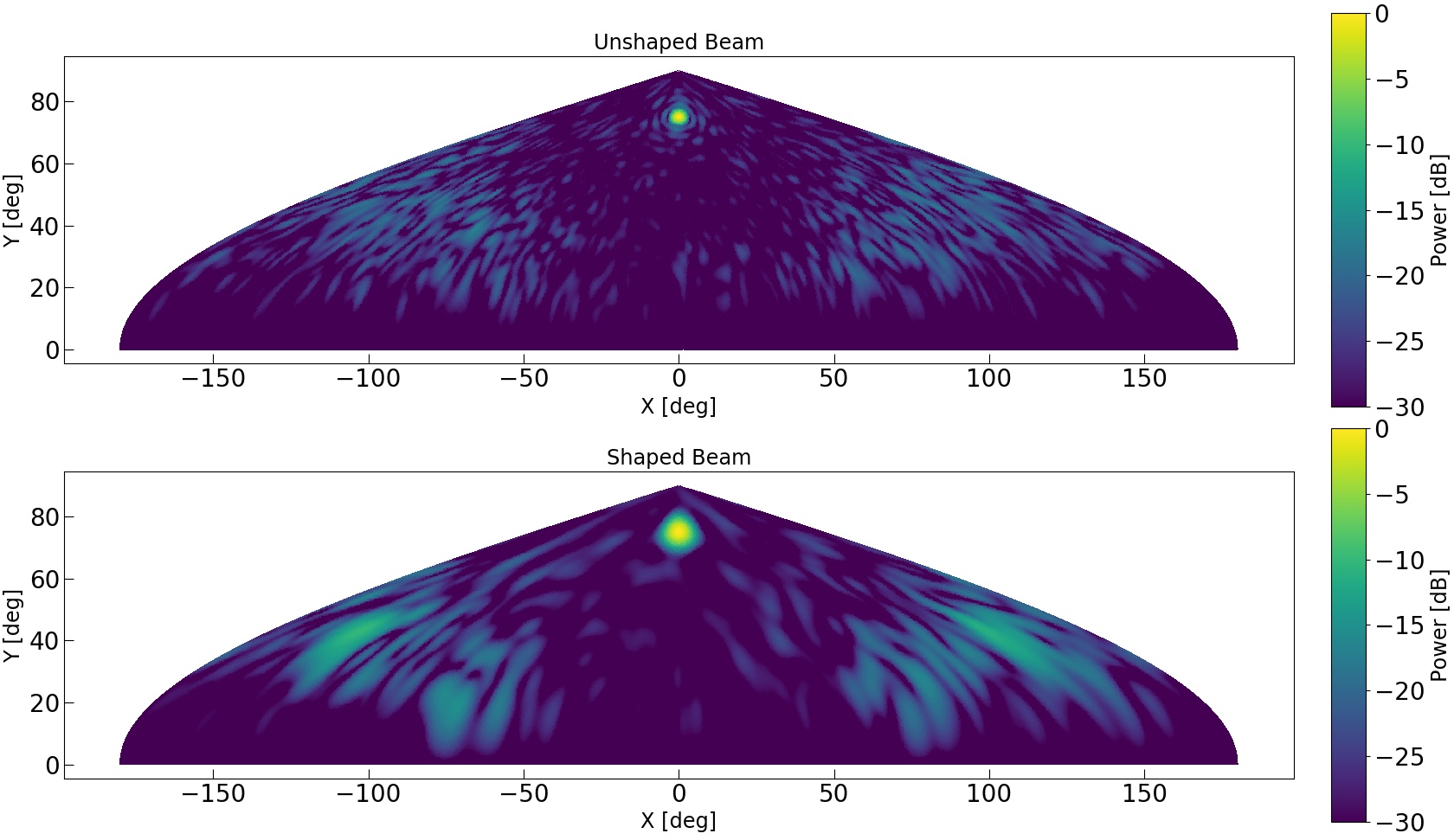}
    \caption{Simulated XX polarization beam patterns for an unshaped (top) and shaped (bottom) beam. Both patterns are 
            for beam pointings at $180\degree$ azimuth and $75\degree$ elevation at $67$ MHz on sinusoidal projection
            of the sky where $X = (a - a_{pointing}) \textrm{cos}(e)$ and $Y = e$
            where $a$ is azimuth, $a_{pointing}$ is the beam pointing
            azimuth, and $e$ is elevation. The bottom plot is a circular beam with a FWHM of $5\degree$ and is a result of the weighting 
            shown in Figure \ref{fig:weights}.}
    \label{fig:beams}
\end{figure}

The weighting scheme shown in Figure \ref{fig:weights} has been computed for a pointing center located at $180\degree$ azimuth 
and $75\degree$ elevation for a tuning at $67$ MHz. The simulated beam pattern for this weighting can be seen at the bottom of
Figure \ref{fig:beams}. It is apparent from the simulations that adjusting the weights to shape the main lobe is feasible, but
the cost is generally stronger side lobes, albeit the side lobes are spatially smoother. This can be an issue if a bright source
enters the side lobes as a low-level Cosmic Dawn signal can be washed out. However, the main lobe becomes more isolated as the
nearby side lobes seem to be pushed outward. This might be beneficial if we can make sure the response from the high side lobes
is minimized by observing during times when no bright sources are in these regions of the sky.

\begin{figure}
    \centering
    \includegraphics[width=\textwidth]{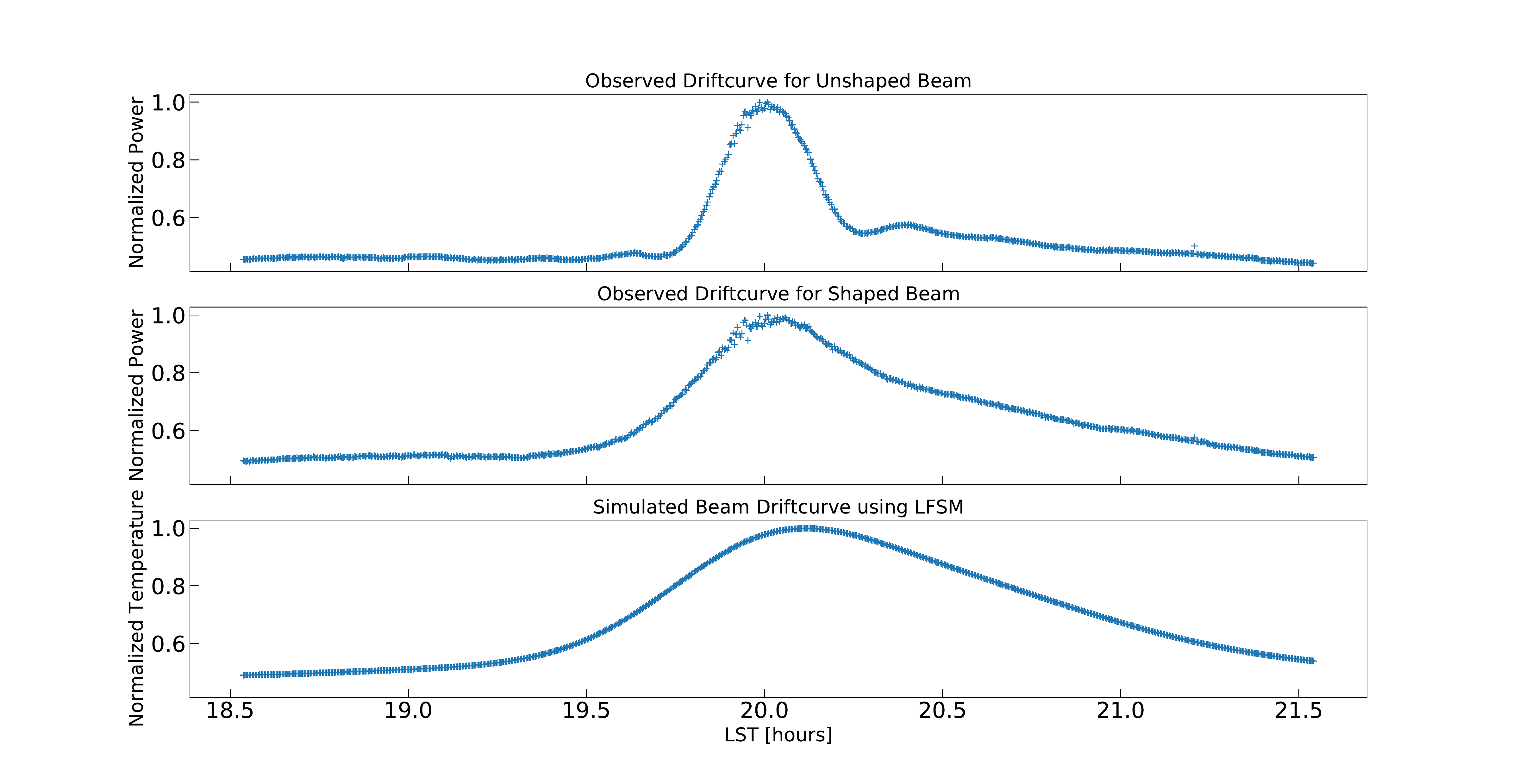}
    \caption{Observed drift curves for an unshaped beam (top) and a custom shaped beam (middle). These drift curves are centered
    around the transit of Cygnus A. The unshaped beam should have a FWHM of $\sim 2.8\degree$ and the custom beam has been made 
    to have a FWHM of $5\degree$. It is apparent that the middle curve corresponds to a wider beam.
    We check the shape against a simulated driftcurve using the LFSM which inherently has $5.1\degree$ resolution. The curves have the
    same general shape implying the shaped beam FWHM is close to $5\degree$.}
    \label{fig:drifts}
\end{figure}

We checked the validity of these simulations by attempting to actually generate a custom beam with FWHM of $5\degree$ at $67$ MHz
pointed at $0\degree$ azimuth and $83.5\degree$ elevation. This pointing allows for Cygnus A to drift through the center of the 
beam at its transit. We pointed the custom beam and collected data for 3 hours centered around the transit of Cygnus A to generate
a drift curve. We can use the shape of the drift curve to test the shape of the beam. The results of this are shown in Figure \ref{fig:drifts}.
We compare the shapes of observed drift curves generated with simultaneous unshaped and shaped beams to a simulated
drift curve generated using a simulation of the shaped beam pattern convolved with the LFSM. All drift curves have been normalized with
respect to the peak value since the observed data is in arbitrary power units and the simulated drift curve has units of Kelvin.
The LFSM has a resolution of $5.1\degree$ and so we look for similarity between the shaped beam drift curve and the simulated one. We 
indeed find that the shaped beam drift curve matches the shape predicted by the LFSM and therefore conclude that the beam is $5\degree$ in
size.

\section{Future Work} \label{future}
The work done thus far has highlighted potential barriers that limit our ability to integrate
to the SNR required for a significant verification of the claims made in \citet{bowman2018}. We believe the
largest contribution to this comes from the beam shape of the array and its chromaticity. The simulations and observations
presented in Section \ref{beamforming} have shown that the beam is shapable, but this does not solve the chromaticity
problem. In theory, this should be straight forward as we can just compute how the dipoles must be weighted for all
frequencies across the band for any given pointing. However, this will be more challenging in practice
since the system will be required to compute and store all the weighting coefficients and apply them
on a per channel basis. Implementing this into LWA-SV is in progress. Shaping the beam size will also allow us to
investigate the performance of different sky models. If we choose a beam size of $> 5\degree$, we will no longer be sub-resolution for
the LWA1 LFSS and GMOSS and can properly test their performance for this application.

We also need to accurately measure the beam and dipole responses and chromaticity. This is essential
if we want to verify that our beam is behaving in a way consistent with the simulations. Consistency is imperative
since the simulations are necessary for our temperature calibration scheme. We have began to experiment with measuring
the dipole radiation pattern using a test drone \citep{chang2015,jacobs2016}. This work is underway, but
this is still a new method and will need to be validated in other ways.

\section{Summary}
We have presented first limits from LWA-SV which employs a new method for attempting to detect the global 
21-cm absorption profile associated with Cosmic Dawn. These limits are still 2 orders of magnitude above what
is required to validate the potential detection reported by EDGES in \citet{bowman2018}. We have detailed
the stability of the system and highlighted the capabilities of a beamforming array as opposed to a single
element radiometer which observes the entire sky. Simultaneous beams allow for \textit{in situ} calibration
of the system which should allow for temperature calibration which is robust against systematics such
as instrumental response and ionospheric effects.

We have selected a large cold region on the sky as our Science Field and use Virgo A as a temperature 
calibrator. We use LWA-SV's spectrometer mode to obtain data products for the XX and YY polarizations. 
We are able to calibrate the observed spectrum of the Science Field and find residual 
r.m.s. limits of $\approx10 \ \rm K$ for XX polarization and $\approx20 \ \rm K$ for YY polarization after fitting
a simple power law to model foreground structure.

The largest challenge that we face is the chromaticity and directionality of the dipole response and the beam. 
We have presented theoretical simulations and early work showing that a direction independent beam can be made
and have outlined what should be needed to move forward to an achromatic beam.

\acknowledgments
Construction of the LWA has been supported by the Office of Naval Research under Contract N00014-07-C-0147 and by 
the AFOSR. Support for operations and continuing development of the LWA is provided by the Air Force Research 
Laboratory and the National Science Foundation under grants AST-1835400 and AGS-1708855.

\facility{LWA}

\newpage
\bibliographystyle{aasjournal}
\bibliography{References.bib}

\end{document}